\documentclass[10pt,a4paper,english,twocolumn]{IEEEtran}
\usepackage[T1]{fontenc}
\usepackage[latin9]{inputenc}
\usepackage{color}
\usepackage{babel}
\usepackage{verbatim}
\usepackage{float}
\usepackage{amsthm}
\usepackage{amsmath}
\usepackage{amssymb}
\usepackage{graphicx}
\usepackage{setspace}
\usepackage{esint}
\makeatletter

\pdfpageheight\paperheight
\pdfpagewidth\paperwidth

\floatstyle{ruled}
\newfloat{algorithm}{tbp}{loa}
\providecommand{\algorithmname}{Algorithm}
\floatname{algorithm}{\protect\algorithmname}

\theoremstyle{plain}
\newtheorem{thm}{\protect\theoremname}
\theoremstyle{plain}
\newtheorem{cor}[thm]{\protect\corollaryname}

\usepackage{subfigure}
\usepackage{epstopdf}
\usepackage{cite}
\usepackage{citesort}
\usepackage{balance}

\makeatother

\providecommand{\corollaryname}{Corollary}
\providecommand{\theoremname}{Theorem}

\begin{document}
% paper title

\title{DNA-GA: A New Approach of Network Performance Analysis% <-this % stops a space
}

\begin{singlespace}

\author{\noindent {\normalsize{}Ming Ding, }\emph{\normalsize{}Data 61, Australia}{\normalsize{}
\{$\mathtt{Ming.Ding@nicta.com.au}$\}}\\
{\normalsize{}David L$\acute{\textrm{o}}$pez P$\acute{\textrm{e}}$rez,
}\emph{\normalsize{}Bell Labs Alcatel-Lucent, Ireland}{\normalsize{}
\{$\mathtt{dr.david.lopez@ieee.org}$\}}\\
{\normalsize{}Guoqiang Mao, }\emph{\normalsize{}The University of
Technology Sydney and Data 61, Australia}{\normalsize{} \{$\mathtt{Guoqiang.Mao@uts.edu.au}$\}}\\
{\normalsize{}Zihuai Lin, }\emph{\normalsize{}The University of Sydney,
Australia}{\normalsize{} \{$\mathtt{zihuai.lin@sydney.edu.au}$\}}}
\end{singlespace}

\maketitle

%\thispagestyle{empty}

% The paper headers
{}
\begin{abstract}
\begin{comment}
In this paper, we analytically derive an upper bound on the error
in approximating the uplink (UL) single-cell interference by a lognormal
distribution in frequency division multiple access (FDMA) small cell
networks (SCNs). Such an upper bound is measured by the Kolmogorov\textendash Smirnov
(KS) distance between the actual cumulative density function (CDF)
and the approximate CDF. The lognormal approximation is important
because it allows tractable network performance analysis. Our results
are more general than the existing works in the sense that we do not
pose any requirement on (i) the shape and/or size of cell coverage
areas, (ii) the uniformity of user equipment (UE) distribution, and
(iii) the type of multi-path fading. Based on our results, we propose
a new framework to directly and analytically investigate a complex
network with practical deployment of multiple BSs placed at irregular
locations, using a power lognormal approximation of the aggregate
UL interference. The proposed network performance analysis is particularly
useful for the 5th generation (5G) systems with general cell deployment
and UE distribution.
\end{comment}
In this paper, we propose a new approach of network performance analysis,
which is based on our previous works on the deterministic network
analysis using the Gaussian approximation (DNA-GA). First, we extend
our previous works to a signal-to-interference ratio (SIR) analysis,
which makes our DNA-GA analysis a formal microscopic analysis tool.
Second, we show two approaches for upgrading the DNA-GA analysis to
a macroscopic analysis tool. Finally, we perform a comparison between
the proposed DNA-GA analysis and the existing macroscopic analysis
based on stochastic geometry. Our results show that the DNA-GA analysis
possesses a few special features: (i) shadow fading is naturally considered
in the DNA-GA analysis; (ii) the DNA-GA analysis can handle non-uniform
user distributions and any type of multi-path fading; (iii) the shape
and/or the size of cell coverage areas in the DNA-GA analysis can
be made arbitrary for the treatment of hotspot network scenarios.
Thus, DNA-GA analysis is very useful for the network performance analysis
of the 5th generation (5G) systems with general cell deployment and
user distribution, both on a microscopic level and on a macroscopic
level.
\footnote{1536-1276 © 2015 IEEE. Personal use is permitted, but republication/redistribution requires IEEE permission. Please find the final version in IEEE from the link: http://ieeexplore.ieee.org/document/7511618/. Digital Object Identifier: 10.1109/ICC.2016.7511618}
\end{abstract}

\section{Introduction}

Due to their potential for large performance gains, dense orthogonal
deployments of small cell networks (SCNs) within the existing macrocell
networks%
\footnote{Orthogonal deployment means that small cells and macrocells operating
on different frequency spectrum, i.e., Small Cell Scenario \#2a defined
in~\cite{TR36.872}.%
} gained much momentum in the design of the 4th generation (4G) systems~\cite{TR36.872},
and are envisaged as the workhorse for capacity enhancement in the
5th generation (5G) systems~\cite{Tutor_smallcell}%
\begin{comment}
, arising from its weak interaction with the macrocell tier, e.g.,
no inter-tier interference
\end{comment}
. In this context, new and more powerful network performance analysis
tools are needed to better understand the performance implications
that these dense orthogonal SCNs bring about~{[}3-10{]}.

Network performance analysis tools can be broadly classified into
two groups, i.e., macroscopic analysis~{[}3,4{]} and microscopic
analysis~{[}5-10{]}. %
\begin{comment}
Generally speaking, the macroscopic analysis gives a general picture
of the network performance, while the microscopic analysis gives more
targeted results for specific networks than the macroscopic analysis.
\end{comment}
The macroscopic analysis usually assumes that user equipments (UEs)
and/or base stations (BSs) are randomly deployed, often following
a homogeneous Poisson%
\begin{comment}
 or uniform distributions
\end{comment}
{} distribution to invoke the stochastic geometry theory~{[}3,4{]}.
In essence, the macroscopic analysis investigates network performance
at a high level, such as coverage probability and signal-to-interference
ratio (SIR) distribution, by averaging over all possible UE and BS
deployments~{[}3,4{]}. Instead, the microscopic analysis allows for
a more detailed analysis and is often conducted assuming that UEs
are randomly placed but that BS locations are known~{[}5-10{]}. Generally
speaking, the macroscopic analysis predicts the network performance
in a statistical sense, while the microscopic analysis is useful for
a network-specific study and optimization%
\begin{comment}
, e.g., optimizing the parameters of UL power control~\cite{UL_interf_sim1}
and performing per-cell loading balance in a specific SCN~\cite{UL_interf_sim2}.
\end{comment}
.

Within the microscopic analysis and paying special attention to uplink
(UL), in~\cite{UL_interf_2cells_ICC}, the authors considered a single
UL interfering cell with a disk-shaped coverage area and presented
closed-form expressions for the UL interference considering both path
loss and shadow fading. In~\cite{UL_interf_LN_conjecture_WCL}, the
authors conjectured that the UL interference in a hexagonal grid based
cellular network may follow a lognormal distribution, which was verified
via simulation. In~\cite{Our_DNA_work_GC15} and~\cite{Our_DNA_work_TWC15},
we went a step further and analytically derived %
\begin{comment}
for the first time
\end{comment}
an upper bound of the error in approximating the dB-scale UL interference
from a single cell by a Gaussian distribution%
\begin{comment}
, while considering any coverage shape and size
\end{comment}
. Such error was measured by the Kolmogorov\textendash Smirnov (KS)
distance~\cite{KS-distance} between the real cumulative density
function (CDF) and the approximate CDF, and it was shown to be small
for practical SCNs. On the basis of this single-cell interference
analysis, we further investigated the approximate distribution of
the aggregate UL interference in a multi-cell scenario as a power
lognormal distribution. For more practical networks, in~\cite{UL_interf_sim1}
and~\cite{UL_interf_sim3}, we also investigated the network performance
of SCNs in current 4G networks using system-level simulations.

In this paper, %
\begin{comment}
In this paper, we focus on the microscopic analysis.
\end{comment}
\begin{comment}
In particular, we consider an UL frequency division multiple access
(FDMA) SCN, which has been widely adopted in the 4th generation (4G)
networks, i.e., the UL single-carrier FDMA (SC-FDMA) system in the
3rd Generation Partnership Project (3GPP) Long Term Evolution (LTE)
networks~\cite{TS36.213} and the UL orthogonal FDMA (OFDMA) system
in the Worldwide Interoperability for Microwave Access (WiMAX) networks~\cite{WiMAX_AI}.
\end{comment}
our objective is to extend the previous works in~\cite{Our_DNA_work_GC15}
and~\cite{Our_DNA_work_TWC15} to analyze the UL SIR performance,
and create a novel and compelling approach for network performance
analysis that can unify the macroscopic and the microscopic analyses
within a single framework, while overcoming a few drawbacks of the
current tools. To that end, our work is composed of the following
three steps:
\begin{enumerate}
\item The extension of the UL interference analysis in~\cite{Our_DNA_work_GC15}
and~\cite{Our_DNA_work_TWC15} to the UL SIR analysis, which makes
our analysis a formal microscopic analysis tool.
\item The upgrade of the developed microscopic analysis tool to a macroscopic
analysis tool.
\item The comparison between the proposed macroscopic analysis tool and
the existing macroscopic analysis based on stochastic geometry.
\end{enumerate}

Since the macroscopic and the microscopic analyses are unified in
our framework based on a deterministic network analysis (DNA) using
the Gaussian approximation (GA) presented in~\cite{Our_DNA_work_GC15}
and~\cite{Our_DNA_work_TWC15}, our framework will be referred to
as the DNA-GA analysis hereafter. As a result of our three-step work,
the contributions of this paper are three-fold, and are summarized
as follows:
\begin{enumerate}
\item Based on the Gaussian approximation theorem presented in~\cite{Our_DNA_work_GC15}
and~\cite{Our_DNA_work_TWC15}, the approximate distributions of
the UL signal power and the UL SIR for the interested UE are derived
in tractable expressions using the Gauss-Hermite numerical integration~\cite{GH_num_integration},
giving rise to the DNA-GA analysis.
\item Although the DNA-GA analysis stands alone as a solid contribution
to the family of microscopic analysis, two approaches for upgrading
the DNA-GA analysis to a macroscopic analysis are further investigated.
The first one is \emph{the semi-analytical }approach, which directly
averages the performance given by many DNA-GA analyses over many random
BS deployments to obtain the performance of the macroscopic analysis.
The second one is \emph{the analytical }approach, which constructs
an idealistic and deterministic BS deployment, and then conducts the
DNA-GA analysis on such BS deployment to obtain an upper-bound performance
of the macroscopic analysis.
\item Interesting results on the comparison between the DNA-GA analysis
and the stochastic geometry analysis~\cite{Jeff_UL} are presented.
Our results show that the DNA-GA analysis qualifies as a new network
performance analysis tool with a few special merits over stochastic
geometry: (i) Shadow fading is naturally considered in the DNA-GA
analysis, while stochastic geometry usually cannot; (ii) Non-uniform
UE distributions and any type of multi-path fading can be treated
in the DNA-GA analysis, while stochastic geometry usually cannot;
(iii) Apart from the common assumption on the cell coverage areas
as Voronoi cells made by stochastic geometry, the shape and/or the
size of cell coverage areas in the DNA-GA analysis can be made arbitrary,
making it suitable for the network performance analysis of hotspot
SCNs%
\begin{comment}
; Thus, the proposed DNA-GA analysis is more general and useful for
network performance analysis
\end{comment}
.
\end{enumerate}

The remainder of the paper is structured as follows. In Section~\ref{sec:Network-Model},
the network scenario and the system model are described. In Section~\ref{sec:DNA-GA-Analysis},
the DNA-GA analysis is presented, followed by its upgrade to a macroscopic
analysis tool in Section~\ref{sec:Macroscopic-Upgrade}. Our results
are validated and compared with those of the stochastic geometry analysis
via simulations in Section~\ref{sec:Simulaiton-and-Discussion}.
Finally, the conclusions are drawn in Section~\ref{sec:Conclusion}.

\section{Network Scenario and System Model\label{sec:Network-Model}}

In this paper, we consider UL transmissions, and assume that each
small cell BS only schedules \emph{one} UE in each frequency/time
resource, i.e., resource block (RB). This is a reasonable assumption
in line with 4G networks, i.e., Long Term Evolution (LTE)~\cite{TS36.213}
and Worldwide Interoperability for Microwave Access (WiMAX)~\cite{WiMAX_AI}.
Note that small cell BSs serving no UE do not contribute to the UL
interference, thereby those BSs are ignored in the analysis.

\begin{comment}
\% network scenario
\end{comment}

\begin{figure}[h]
\vspace{-0.3cm}

\noindent \begin{centering}
\includegraphics[width=6cm]{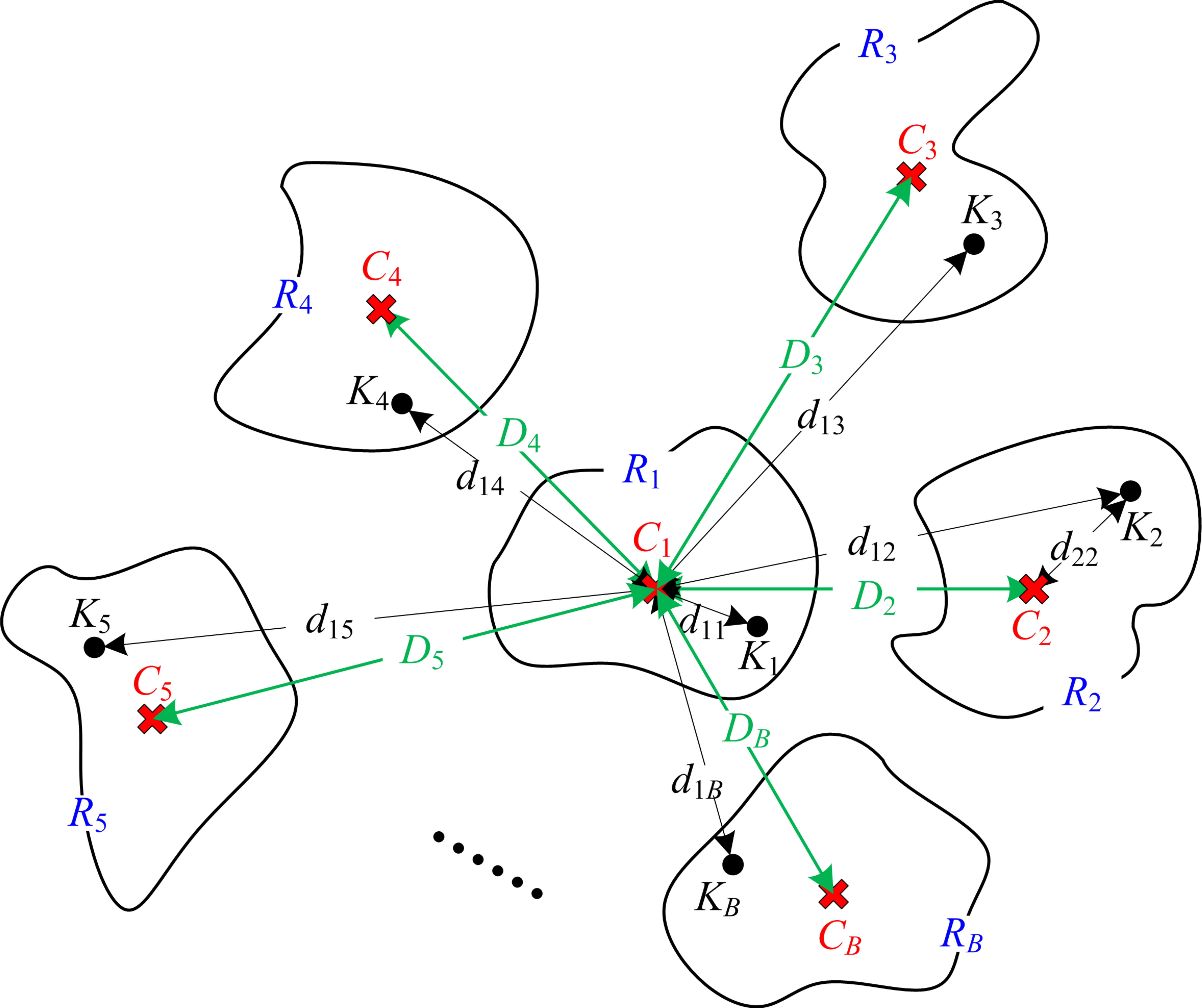} \renewcommand{\figurename}{Fig.}\protect\caption{\label{fig:sys_model}A schematic model of the considered SCN.}

\par\end{centering}

\vspace{-0.3cm}
\end{figure}

Regarding the network scenario, we consider a SCN with multiple small
cells operating on the same carrier frequency, as shown in Fig.~\ref{fig:sys_model}.
In more detail, the SCN consists of $B$ small cells, each of which
is managed by a BS. The network includes the small cell of interest
denoted by $C_{1}$ and $B-1$ interfering small cells denoted by
$C_{b},b\in\left\{ 2,\ldots,B\right\} $. We focus on a particular
RB, and denote by $K_{b}$ the active UE associated with small cell
$C_{b}$ in such RB. Moreover, we denote by $R_{b}$\emph{ the coverage
area} of small cell $C_{b}$, in which its associated UEs are randomly
distributed. Note that the coverage areas of adjacent small cells
may overlap due to the arbitrary shapes and sizes of $\left\{ R_{b}\right\} ,b\in\left\{ 2,\ldots,B\right\} $.

\begin{comment}
\% a deterministic deployment and a random drop scenario
\end{comment}

\textit{The distance} (in km) from the BS of $C_{b}$ to the BS of
$C_{1}$, $b\in\left\{ 1,\ldots,B\right\} $, and the distance from
UE $K_{b}$ to the BS of $C_{m}$, $b,m\in\left\{ 1,\ldots,B\right\} $,
are denoted by $D_{b}$ and $d_{bm}$, respectively. Since DNA-GA
is a microscopic analysis tool, we consider a deterministic deployment
of BSs, i.e., the set $\left\{ D_{b}\right\} $ is known, while each
UE $K_{b}$ is randomly distributed in $R_{b}$ with a distribution
function $f_{Z_{b}}\left(z\right),z\in R_{b}$. Hence, $d_{bm}$ is
a random variable (RV), whose distribution cannot be readily expressed
in an analytical form due to the arbitrary shape and size of $R_{b}$,
and the arbitrary form of $f_{Z_{b}}\left(z\right)$. Regarding $R_{b}$
and $f_{Z_{b}}\left(z\right)$, we have two remarks in the following.

\noindent \textbf{Remark 1:} Unlike the existing works, e.g.,~{[}3-6,
9-10{]}, where only the uniform UE distribution was considered, DNA-GA
can handle any probability density function (PDF) of general UE distribution,
here denoted by $f_{Z_{b}}\left(z\right)$, where $0<f_{Z_{b}}\left(z\right)<+\infty,z\in R_{b}$
and its integral over $R_{b}$ equals to one, i.e., $\int_{R_{b}}f_{Z_{b}}\left(z\right)dz=1$.

\noindent \textbf{Remark 2:} Even if $f_{Z_{b}}(z)$ is constant with
$z$, we can only say that the UE distribution is uniform within the
small cell coverage area $R_{b}$, but we cannot guarantee that the
UE distribution is uniform within the entire scenario, because no
UEs are deployed outside the hotspot areas $\left\{ R_{b}\right\} $,
which may cause the non-uniformity of UE distribution within the entire
scenario. Instead, in stochastic geometry~{[}3,4{]}, UEs are usually
assumed to be uniformly distributed within the entire scenario, creating
Voronoi cells, which is less general and practical than our assumption
of $R_{b}$ and $f_{Z_{b}}\left(z\right)$. Note that in the sequel,
the characterization of UE distribution is meant within $R_{b}$.

\begin{comment}
\begin{table*}[!tp]
\begin{centering}
{\small{}\protect\caption{\label{tab:RV_def}Definition of RVs.}
}
\par\end{centering}{\small \par}

{\small{}\vspace{-0.2cm}
}{\small \par}

\begin{centering}
{\small{}}%
\begin{tabular}{|l|l|l|}
\hline
{\small{}RV} & {\small{}Description} & {\small{}Distribution}\tabularnewline
\hline
\hline
{\small{}$Z_{b}$} & {\small{}The position of $K_{b}$ in $R_{b}$} & {\small{}The PDF is $f_{Z_{b}}\left(z\right),z\in R_{b}$}\tabularnewline
\hline
{\small{}$d_{bm}$} & {\small{}The distance (in km) from $K_{b}$ to $C_{m}$} & {\small{}Related to $f_{Z_{b}}\left(z\right)$, $R_{b}$ and $D_{b}$}\tabularnewline
\hline
{\small{}$L_{bm}$} & {\small{}The path loss (in dB) from $K_{b}$ to $C_{m}$} & {\small{}Related to $f_{Z_{b}}\left(z\right)$, $R_{b}$ and $D_{b}$}\tabularnewline
\hline
{\small{}$S_{bm}$} & {\small{}The shadow fading (in dB) from $K_{b}$ to $C_{m}$} & {\small{}i.i.d. $\mathcal{N}\left(0,\sigma_{\textrm{Shad}}^{2}\right)$}\tabularnewline
\hline
{\small{}$H_{bm}$} & {\small{}The channel gain (in dB) from $K_{b}$ to $C_{m}$} & {\small{}i.i.d. distribution with the PDF of $f_{H}\left(h\right)$}\tabularnewline
\hline
{\small{}$P_{b}$} & {\small{}The UL transmission power (in dBm) of $K_{b}$} & {\small{}Related to $f_{Z_{b}}\left(z\right)$, $R_{b}$ and $S_{bb}$}\tabularnewline
\hline
\end{tabular}
\par\end{centering}{\small \par}

\vspace{0.2cm}

\rule[0.5ex]{2\columnwidth}{1pt}
\end{table*}
\end{comment}

Next, we present the modeling of path loss, shadow fading, UL transmission
power, %
\begin{comment}
reception filter
\end{comment}
multi-path fading and noise%
\begin{comment}
, respectively
\end{comment}
.

\begin{comment}
\% path loss
\end{comment}

Based on the definition of $d_{bm}$, \textit{the path loss} in dB
from UE $K_{b}$ to the BS of $C_{m}$ is modeled as

\begin{singlespace}
\noindent
\begin{equation}
L_{bm}={A}+{\alpha}\times10{\log_{10}}{d_{bm}},\label{eq:PL_BS2UE}
\end{equation}

\end{singlespace}

\noindent where $A$ is the path loss in dB at the reference distance
of $d_{bm}=1$ and $\alpha$ is the path loss exponent. In practice,
$A$ and $\alpha$ are constants obtainable from field tests~\cite{TR36.828}.
Note that $L_{bm}$ is a RV due to the randomness of $d_{bm}$.

\noindent %
\begin{comment}
\noindent Moreover, reciprocity is assumed for path losses, i.e.,
\textit{\emph{the path loss}}\emph{ }from UE $K_{m}$ to the BS of
$C_{b}$ equals to $L_{bm}$.
\end{comment}

\noindent %
\begin{comment}
{[}Ming{]}: In this paper, we will not touch the LoS/NLoS thing.

Deleted words for future use:

\noindent It is important to note that for a certain path loss model,
$A_{1}$ and $\alpha_{1}$ may take different values for the line-of-sight
(LoS) transmission case and the non-LoS (NLoS) transmission case.
Besides, the LoS probability is normally a monotonically decreasing
function with respect to the distance $dis$ in km between the transmitter
and the receiver. For example, in~\cite{TR36.828}, the LoS probability
function is modeled as

\begin{singlespace}
\noindent
\begin{eqnarray}
\textrm{Pr}^{\textrm{LoS}}\left(dis\right) & = & 0.5-\min\left\{ {0.5,5\exp\left({-0.156/dis}\right)}\right\} \nonumber \\
 &  & +\min\left\{ {0.5,5\exp\left({-1\times dis/0.03}\right)}\right\} .\label{eq:LoS_Prob_fun}
\end{eqnarray}
\end{singlespace}
\end{comment}

\noindent %
\begin{comment}
For the convenience of mathematical expression in the sequel, the
linear-scale value of $L_{bm}$ is defined as $\gamma_{bm}={10}^{\frac{{L_{bm}}}{{10}}}$,
which is a RV due to the randomness of $L_{bm}$($d_{bm}$).
\end{comment}

\begin{singlespace}
\noindent %
\begin{comment}
\begin{singlespace}
\noindent
\begin{equation}
\gamma_{bm}={10}^{\frac{{L_{bm}}}{{10}}},\label{eq:aux_gamma}
\end{equation}
\end{singlespace}
\end{comment}

\end{singlespace}

\begin{comment}
\% shadow fading
\end{comment}

\textit{The shadow fading} in dB from UE $K_{b}$ to the BS of $C_{m}$
is denoted by $S_{bm}$%
\begin{comment}
$\left(b,m\in\left\{ 1,\ldots,B\right\} \right)$
\end{comment}
, and is usually assumed to follow a lognormal distribution~\cite{TR36.828}.
Based on this assumption, $S_{bm}$ is modeled as an independently
and identically distributed (i.i.d.) zero-mean Gaussian RV with a
variance of $\sigma_{\textrm{Shad}}^{2}$, i.e., $S_{bm}\sim\mathcal{N}\left(0,\sigma_{\textrm{Shad}}^{2}\right)$%
\begin{comment}
\noindent Note that a more realistic assumption would be the correlated
shadow fading~\cite{Corr_shadow_fading}, which constructs $S_{bm}$
and $S_{jm}$ $\left(b,j,m\in\left\{ 1,\ldots,B\right\} ,\, b\neq j\right)$
as correlated RVs, where the correlation coefficient decreases with
the increase of the distance from UE $K_{b}$ to UE $K_{j}$. Such
assumption of the correlated shadow fading complicates the analysis
since it is difficult to characterize the distribution of the inter-UE
distance. For the sake of tractability, in this paper, we assume i.i.d.
shadow fading for the UE-to-BS links.
\end{comment}
.

\textit{The UL transmission power} in dBm of UE $K_{b}$ is denoted
by $P_{b}$, and is subject to a semi-static power control (PC) mechanism%
\begin{comment}
\noindent Note that in practice $P_{b}$ is also constrained by the
maximum value of the UL power, denoted by $P_{\textrm{max}}$ at the
UE. However, the power constraint is a minor issue for UEs in SCNs
since they are generally not power-limited due to the close proximity
of a UE and its associated small cell BS. For example, it is recommended
in~\cite{TR36.828} that $P_{\textrm{max}}$ is smaller than the
small cell BS DL power by only 1dB, which grants a similar coverage
range for the DL and the UL. Therefore, the UL power limitation is
a minor issue as long as the UE is able to connect with the BS in
the DL. For the sake of tractability, we model $P_{b}$ as (\ref{eq:UL_P_UE}),
which has been widely adopted in the literature {[}4-8{]}.
\end{comment}
, e.g., the fractional path loss compensation (FPC) scheme~\cite{TR36.828}.
Based on this FPC scheme, $P_{b}$ %
\begin{comment}
\noindent  and affected by the per-UE signal-to-interference-plus-noise
ratio (SINR) target
\end{comment}
\begin{comment}
\noindent On the other hand, the per-UE SINR target complicates the
analysis since it is difficult to model the distribution of the target
SINRs.
\end{comment}
is modeled as%
\begin{comment}
An important note is about the power constraint of UEs. According
to~\cite{TR36.828}, $P_{b}$ should not be larger than 23\ dBm.
Such constraint can be readily satisfied in SCNs since For example,
for a distance of 100m and a large shadow fading of 30dB, there still
exists a power headroom of 20dB.

missing that the headroom also depends on
\end{comment}

\noindent
\begin{equation}
P_{b}={P_{0}}+\eta\left({L_{bb}+S_{bb}}\right),\label{eq:UL_P_UE}
\end{equation}
where $P_{0}$ is the target received power in dBm on the considered
RB at the BS, $\eta\in\left(0,1\right]$ is the FPC factor, and $L_{bb}$
and $S_{bb}\sim\mathcal{N}\left(0,\sigma_{\textrm{Shad}}^{2}\right)$
have been discussed above.

\begin{comment}
{[}Ming{]}: For future use:

\textit{The multi-path fading} channel from the BS of $C_{b}$ to
UE $K_{m}$ is denoted by ${\bf {h}}_{bm}^{{\rm {BS2UE}}}\in\mathbb{C}^{1\times N}$.
All channels are assumed to experience uncorrelated flat Rayleigh
fading on the considered resource block, and the channel coefficients
are modeled as i.i.d. zero-mean circularly symmetric complex Gaussian
(ZMCSCG) RVs with unit variance. Note that reciprocity is assumed
for the multi-path fading from the BS of $C_{b}$ to UE $K_{m}$ and
the multi-path fading from UE $K_{m}$ to the BS of $C_{b}$, i.e.,
it is always true that ${\bf {h}}_{mb}^{{\rm {UE2BS}}}=\left({\bf {h}}_{bm}^{{\rm {BS2UE}}}\right)^{\textrm{H}}$.
\end{comment}

\begin{comment}
\% channel modeling (an UL FDMA system with 1 by 1 SISO)
\end{comment}

\textit{The multi-path fading} channel %
\begin{comment}
vector
\end{comment}
from UE $K_{b}$ to the BS of $C_{m}$ is denoted by ${\bf {h}}_{bm}\in\mathbb{C}$%
\begin{comment}
${\bf {h}}_{bm}\in\mathbb{C}^{N\times1}$
\end{comment}
, where we assume that each UE and each BS are equipped with one omni-directional
antenna. It is important to note that we consider a general type of
multi-path fading by assuming that the effective channel gain in dB
associated with ${\bf {h}}_{bm}$ is defined as $H_{bm}=10\log_{10}\left|{\bf {h}}_{bm}\right|^{2}$,
which follows an i.i.d. distribution with a PDF of $f_{H}\left(h\right)$.
For example, $\left|{\bf {h}}_{bm}\right|^{2}$ can be characterized
by an exponential distribution or a Gamma distribution in case of
Rayleigh fading or Nakagami fading, respectively~\cite{Book_Proakis}.
And hence, the distribution of $H_{bm}$ can be derived analytically%
\begin{comment}
 according to the distribution of $\left|{\bf {h}}_{bm}\right|^{2}$
\end{comment}
.

\noindent %

\noindent %
\begin{comment}
\% reception filter
\end{comment}
\emph{}%

\noindent %

Finally, we ignore the additive \emph{noise} because the 4G and the
5G SCNs generally work in the interference-limited region~\cite{Tutor_smallcell}.
\begin{comment}
For clarity, the defined RVs in our system model are summarized in
Table~\ref{tab:RV_def}.
\end{comment}

\section{The Proposed DNA-GA Analysis\label{sec:DNA-GA-Analysis}}

\begin{comment}
Placeholder
\end{comment}

The proposed DNA-GA analysis consists of three steps, i.e., the interference
analysis, the signal power analysis, and the SIR analysis, which are
presented in the following.

\subsection{The Interference Analysis}

\begin{comment}
Placeholder
\end{comment}

Based on the definition of RVs discussed in Section~\ref{sec:Network-Model}%
\begin{comment}
listed in Table~\ref{tab:RV_def}
\end{comment}
, \emph{the UL received interference power} in dBm from UE $K_{b}$
to the BS of $C_{1}$ can be written as%

{\small{}\vspace{-0.1cm}
}{\small \par}

\noindent
\begin{eqnarray}
I_{b} & \hspace{-0.3cm}\overset{(a)}{=}\hspace{-0.3cm} & P_{b}-L_{b1}-S_{b1}+H_{b1}\nonumber \\
 & \hspace{-0.3cm}=\hspace{-0.3cm} & P_{0}+\left(\eta L_{bb}-L_{b1}\right)+\left(\eta S_{bb}-S_{b1}\right)+H_{b1}\nonumber \\
 & \hspace{-0.3cm}\overset{\triangle}{=}\hspace{-0.3cm} & \left(P_{0}+L+S\right)+H_{b1},\nonumber \\
 & \hspace{-0.3cm}\overset{\triangle}{=}\hspace{-0.3cm} & I_{b}^{\left(1\right)}+H_{b1},\label{eq:rx_interf_I1b_UL}
\end{eqnarray}

\noindent where (\ref{eq:UL_P_UE}) is plugged into the step (a) of
(\ref{eq:rx_interf_I1b_UL}), and $L$ and $S$ are defined as $L\overset{\triangle}{=}\left(\eta L_{bb}-L_{b1}\right)$
and $S\overset{\triangle}{=}\left(\eta S_{bb}-S_{b1}\right)$, respectively.
Apparently, $L$ and $S$ are independent RVs. Besides, the first
part of $I_{b}$ is further defined as $I_{b}^{\left(1\right)}\overset{\triangle}{=}\left(P_{0}+L+S\right)$.
\begin{comment}
\noindent In~(), according to~(), $\xi_{1b}^{-1}\tilde{\xi}_{bb}$
can be represented by another lognormal RV as

\noindent
\begin{eqnarray}
\xi_{1b}^{-1}\tilde{\xi}_{bb} & \hspace{-0.3cm}=\hspace{-0.3cm} & 10^{-\frac{1}{10}S_{1b}}10^{\frac{1}{{10}}\left({\eta S_{bb}}\right)}\nonumber \\
 & \hspace{-0.3cm}=\hspace{-0.3cm} & 10^{-\frac{1}{{10}}\left(\sqrt{\rho}S_{b}^{{\rm {UE}}}+\sqrt{1-\rho}S_{1}^{{\rm {BS}}}\right)}10^{\frac{1}{{10}}\left(\eta\left(\sqrt{\rho}S_{b}^{{\rm {UE}}}+\sqrt{1-\rho}S_{b}^{{\rm {BS}}}\right)\right)}\nonumber \\
 & \hspace{-0.3cm}=\hspace{-0.3cm} & 10^{\frac{1}{{10}}\left(\eta\sqrt{1-\rho}S_{b}^{{\rm {BS}}}-\left(1-\eta\right)\sqrt{\rho}S_{b}^{{\rm {UE}}}-\sqrt{1-\rho}S_{1}^{{\rm {BS}}}\right)}\nonumber \\
 & \hspace{-0.3cm}\overset{\triangle}{=}\hspace{-0.3cm} & \xi_{I_{1b}}.\label{eq:comb_lognormal_RV}
\end{eqnarray}

\noindent For convenience, we define the dB scale RV associated with
$\xi_{I_{1b}}$ as
\end{comment}
\begin{comment}
\begin{singlespace}
\noindent
\begin{eqnarray}
S & \hspace{-0.3cm}=\hspace{-0.3cm} & \eta\sqrt{1-\rho}S_{b}^{{\rm {BS}}}-\left(1-\eta\right)\sqrt{\rho}S_{b}^{{\rm {UE}}}-\sqrt{1-\rho}S_{1}^{{\rm {BS}}}.\hspace{0.8cm}
\end{eqnarray}
\end{singlespace}
\end{comment}
Since $S_{bb}$ and $S_{b1}$ $\left(b\in\left\{ 2,\ldots,B\right\} \right)$
are i.i.d. zero-mean Gaussian RVs, it is easy to show that $S$ is
also a Gaussian RV, whose mean and variance are

\noindent
\begin{equation}
\begin{cases}
\mu_{S}=0 & \hspace{-0.3cm}\\
\sigma_{S}^{2}=\left(1+\eta^{2}\right)\sigma_{\textrm{Shad}}^{2} & \hspace{-0.3cm}
\end{cases}.\label{eq:comb_shadowing_mean_and_var}
\end{equation}

From the definition of \emph{$I_{b}$} in~(\ref{eq:rx_interf_I1b_UL}),\emph{
the aggregate interference power} in mW from all interfering UEs to
the BS of $C_{1}$ can be formulated as

\noindent
\begin{equation}
I^{\textrm{mW}}=\sum\limits _{b=2}^{B}{10^{\frac{1}{10}I_{b}}}.\label{eq:rx_interf_UL}
\end{equation}

\begin{comment}
\noindent where in step (a) $\mathbb{E}_{\left[X\right]}\left\{ \cdot\right\} $
denotes the expectation operation taken with respect to $X$, and
step (b) is obtained considering the decomposition of $P_{b}^{{\rm {UE}}}$
in (\ref{eq:decomp_UL_P_UE}) and that $I_{1b}$ is defined as
\end{comment}

In our previous work~\cite{Our_DNA_work_TWC15}, we show that the
distribution of $I^{\textrm{mW}}$ can be well approximated by a power
lognormal distribution. This approximation is summarized in the following.

\subsubsection{The Distribution of $I_{b}^{\left(1\right)}$ in (\ref{eq:rx_interf_I1b_UL})\label{sub:Ib(1)}}

First, we analyze the distribution of $I_{b}^{\left(1\right)}$ shown
in (\ref{eq:rx_interf_I1b_UL}). Considering a small approximation
error, upper-bounded by the KS distance~\cite{KS-distance} provided
in~\cite{Our_DNA_work_TWC15}, we approximate $I_{b}^{\left(1\right)}$
by a Gaussian RV $G_{b}$, whose mean and variance are

\noindent
\begin{equation}
\begin{cases}
\mu_{G_{b}}=P_{0}+\mu_{L}+\mu_{S} & \hspace{-0.3cm}\\
\ensuremath{\sigma_{G_{b}}^{2}}=\sigma_{L}^{2}+\sigma_{S}^{2} & \hspace{-0.3cm}
\end{cases},\label{eq:approx_LxLN_mean_and_var}
\end{equation}

\noindent where $\mu_{L}$ and $\sigma_{L}^{2}$ are respectively
the mean and the variance of $L$, which can be obtained using numerical
integration involving $f_{Z_{b}}\left(z\right)$ and $R_{b}$~{[}7,8{]}.
Details are omitted for brevity.

\subsubsection{The Distribution of $I_{b}$ in (\ref{eq:rx_interf_I1b_UL})\label{sub:Ib}}

Second, we analyze the distribution of $I_{b}=I_{b}^{\left(1\right)}+H_{b1}$
shown in (\ref{eq:rx_interf_I1b_UL}). Considering a small approximation
error, upper-bounded by the KS distance~\cite{KS-distance} provided
in~\cite{Our_DNA_work_TWC15}, we approximate $I_{b}$ by another
Gaussian RV $Q_{b}$, whose mean and variance are

\noindent
\begin{equation}
\begin{cases}
\mu_{Q_{b}}=\mu_{G_{b}}+\mu_{H_{b1}} & \hspace{-0.3cm}\\
\ensuremath{\sigma_{Q_{b}}^{2}}=\sigma_{G_{b}}^{2}+\sigma_{H_{b1}}^{2} & \hspace{-0.3cm}
\end{cases}.\label{eq:approx_LxLNxEXP_mean_and_var}
\end{equation}

\noindent where $\mu_{H_{b1}}$ and $\sigma_{H_{b1}}^{2}$ are respectively
the mean and the variance of $H_{b1}$. We omit the details for brevity.

Note that the upper bound of the total approximation error of the
above two steps is obtained from the summation of the individual approximation
errors of the two steps in closed-form expressions~\cite{Our_DNA_work_TWC15}.
And it has been shown in~\cite{Our_DNA_work_TWC15} that the total
approximation error is small for practical SCNs, without any requirement
on (i) the uniformity of UE distribution and/or the type of multi-path
fading; and (ii) the shape and/or size of cell coverage areas. Intuitively
speaking, the results in~\cite{Our_DNA_work_TWC15} show that the
larger the variance of the Gaussian RV, i.e., $S$ in (\ref{eq:approx_LxLN_mean_and_var})
or $G_{b}$ in (\ref{eq:approx_LxLNxEXP_mean_and_var}), the better
the approximation in (\ref{eq:approx_LxLN_mean_and_var}) or in (\ref{eq:approx_LxLNxEXP_mean_and_var}),
due to the increasing dominance of the Gaussian RV.

\subsubsection{The Distribution of $I^{\textrm{mW}}$ in (\ref{eq:rx_interf_UL})\label{sub:ImW}}

Third, we invoke the main results in~{[}17-18{]}%
\begin{comment}
\cite{power_LN_approx_TVT}-\hspace{-0.03cm}\cite{power_LN_approx_GC}
\end{comment}
, which indicate that the sum of multiple independent lognormal RVs
can be well approximated by a power lognormal RV. Accordingly, in
our case, since each $I_{b},b\in\left\{ 2,\dots,B\right\} $ is approximated
by a Gaussian RV $Q_{b}$, their sum $10^{\frac{1}{{10}}Q_{b}}$ shown
in (\ref{eq:rx_interf_UL}) should be well approximated by a power
lognormal RV expressed as $\hat{I}^{\textrm{mW}}=10^{\frac{1}{{10}}Q}$,
where the PDF and CDF of $Q$~\cite{power_LN_approx_TVT} can be
written as (\ref{eq:PDF_CDF_powerLN_Q}) shown on the top of the next
page. In (\ref{eq:PDF_CDF_powerLN_Q}), $\Phi\left(x\right)$ is the
CDF of the standard normal distribution, and the parameters $\lambda$,
$\mu_{Q}$ and $\sigma_{Q}$ are obtained from $\left\{ \mu_{Q_{b}}\right\} $
and $\left\{ \ensuremath{\sigma_{Q_{b}}^{2}}\right\} $ that are computed
by (\ref{eq:approx_LxLNxEXP_mean_and_var}). The procedure to obtain
$\lambda$, $\mu_{Q}$ and $\sigma_{Q}$ is omitted here for brevity,
but interested readers are referred to Appendix~B of~\cite{Our_DNA_work_TWC15}
for further details. As a result of (\ref{eq:PDF_CDF_powerLN_Q}),
the PDF and CDF of $\hat{I}^{\textrm{mW}}$ can be written as (\ref{eq:PDF_CDF_agg_Interf})
shown on the top of the next page, where $\zeta=\frac{10}{\ln10}$
is a scalar factor originated from the variable change from $10\log_{10}v$
to $\ln v$.

Finally, we approximate the distribution of $I^{\textrm{mW}}$ by
that of $\hat{I}^{\textrm{mW}}$ shown in (\ref{eq:PDF_CDF_agg_Interf})
presented on the top of the next page. Note that in this step, the
approximation error depends on the approximate error introduced by
the power lognormal approximation, which has been shown to be reasonably
small and good enough in practical cases~{[}17-18{]}%
\begin{comment}
\cite{power_LN_approx_TVT}-\hspace{-0.03cm}\cite{power_LN_approx_GC}
\end{comment}
.

\begin{comment}
Analysis of UL SINR
\begin{itemize}
\item Note that the lognormal approximation is NOT very accurate for the
signal power because of the UL power control mechanism. In more detail,
the lognormal shadowing fading is mostly compensated by the UL transmission
power, rendering a less dominant role of the lognormal shadowing fading
compared with the multi-path fading (Chi-squared distribution with
$2N$ degrees of freedom). Therefore, we may have to leave the expression
of the signal power as it is and perform no approximation for it.
On a side note, this is not a problem for the DL analysis, where the
transmission power is constant and the lognormal shadowing fading
is always a major player.
\item Regarding the SINR distribution, ``S'' follows a distribution with
a complicated analytical expression, ``I'' can be approximated by
a power lognormal RV, ``N'' is a constant. I have been working on
the math of the SINR distribution in the past few weeks and I think
I finally know how to do it :) Let me complete this section later. \end{itemize}
\end{comment}

\subsection{The Signal Power Analysis\label{sec:Analysis-of-Signal}}

\begin{comment}
Placeholder
\end{comment}

\begin{algorithm*}
\noindent
\begin{equation}
\begin{cases}
\textrm{PDF of }Q:\:{f_{Q}}\left(q\right)=\lambda\Phi^{\lambda-1}\left(\frac{q-{\mu_{Q}}}{\sigma_{Q}}\right)\frac{1}{\sqrt{2\pi\ensuremath{\sigma_{Q}^{2}}}}\exp\left\{ {-\frac{{\left({q-{\mu_{Q}}}\right)^{2}}}{{2\ensuremath{\sigma_{Q}^{2}}}}}\right\}  & \hspace{-0.3cm}\\
\textrm{CDF of }Q:\:{F_{Q}}\left(q\right)=\Phi^{\lambda}\left(\frac{q-{\mu_{Q}}}{\sigma_{Q}}\right) & \hspace{-0.3cm}
\end{cases}.\label{eq:PDF_CDF_powerLN_Q}
\end{equation}

\noindent
\begin{equation}
\begin{cases}
\textrm{PDF of }\hat{I}^{\textrm{mW}}:\:{f_{\hat{I}^{\textrm{mW}}}}\left(v\right)=\lambda\Phi^{\lambda-1}\left(\frac{\zeta\ln v-{\mu_{Q}}}{\sigma_{Q}}\right)\frac{\zeta}{v\sqrt{2\pi\ensuremath{\sigma_{Q}^{2}}}}\exp\left\{ {-\frac{{\left({\zeta\ln v-{\mu_{Q}}}\right)^{2}}}{{2\ensuremath{\sigma_{Q}^{2}}}}}\right\}  & \hspace{-0.3cm}\\
\textrm{CDF of }\hat{I}^{\textrm{mW}}:\:{F_{\hat{I}^{\textrm{mW}}}}\left(v\right)=\Phi^{\lambda}\left(\frac{\zeta\ln v-{\mu_{Q}}}{\sigma_{Q}}\right) & \hspace{-0.3cm}
\end{cases}.\label{eq:PDF_CDF_agg_Interf}
\end{equation}
\end{algorithm*}

Based on the definition of RVs discussed in Section~\ref{sec:Network-Model},
\emph{the UL received signal power} in dBm from UE $K_{1}$ to the
BS of $C_{1}$ can be written as

{\small{}\vspace{-0.1cm}
}{\small \par}

\noindent
\begin{eqnarray}
X_{1} & \hspace{-0.3cm}\overset{(a)}{=}\hspace{-0.3cm} & P_{1}-L_{11}-S_{11}+H_{11}\nonumber \\
 & \hspace{-0.3cm}=\hspace{-0.3cm} & P_{0}+\left(\eta L_{11}-L_{11}\right)+\left(\eta S_{11}-S_{11}\right)+H_{11}\nonumber \\
 & \hspace{-0.3cm}\overset{\triangle}{=}\hspace{-0.3cm} & \left(P_{0}+\bar{L}_{11}+\bar{S}_{11}\right)+H_{11},\nonumber \\
 & \hspace{-0.3cm}\overset{\triangle}{=}\hspace{-0.3cm} & X_{1}^{\left(1\right)}+H_{11},\label{eq:rx_signal_X11_UL}
\end{eqnarray}

\noindent where (\ref{eq:UL_P_UE}) is plugged into the step (a) of
(\ref{eq:rx_signal_X11_UL}). Besides, $\bar{L}_{11}$ and $\bar{S}_{11}$
are defined as $\bar{L}_{11}\overset{\triangle}{=}\left(\eta-1\right)L_{11}$
and $\bar{S}_{11}\overset{\triangle}{=}\left(\eta-1\right)S_{11}$,
respectively. %
\begin{comment}
\noindent Apparently, $\bar{L}_{11}$ and $\bar{S}_{11}$ are independent
RVs.
\end{comment}
The first part of $X_{1}$ is further defined as $X_{1}^{\left(1\right)}\overset{\triangle}{=}P_{0}+\bar{L}_{11}+\bar{S}_{11}$,
and it is easy to show that $\bar{S}_{11}$ is a Gaussian RV, whose
mean and variance are

\noindent
\begin{equation}
\begin{cases}
\mu_{\bar{S}_{11}}=0 & \hspace{-0.3cm}\\
\sigma_{\bar{S}_{11}}^{2}=\left(1-\eta\right)^{2}\sigma_{\textrm{Shad}}^{2} & \hspace{-0.3cm}
\end{cases}.\label{eq:singal_shadowing_mean_and_var}
\end{equation}

Similar to the discussion in subsection~\ref{sub:Ib(1)}, we consider
a small approximation error, upper-bounded by the KS distance shown
in~\cite{Our_DNA_work_TWC15}, and we approximate $X_{1}^{\left(1\right)}$
by a Gaussian RV $G_{1}$, whose mean and variance are

\noindent
\begin{equation}
\begin{cases}
\mu_{G_{1}}=P_{0}+\mu_{\bar{L}_{11}}+\mu_{\bar{S}_{11}} & \hspace{-0.3cm}\\
\ensuremath{\sigma_{G_{1}}^{2}}=\sigma_{\bar{L}_{11}}^{2}+\sigma_{\bar{S}_{11}}^{2} & \hspace{-0.3cm}
\end{cases},\label{eq:approx_signal_LxLN_mean_and_var}
\end{equation}

\noindent where $\mu_{\bar{L}_{11}}$ and $\sigma_{\bar{L}_{11}}^{2}$
are respectively the mean and the variance of $\bar{L}_{11}$. As
a result, (\ref{eq:rx_signal_X11_UL}) can be re-formulated as

\begin{singlespace}
\noindent
\begin{equation}
X_{1}\approx G_{1}+H_{11}\overset{\triangle}{=}\hat{X}_{1}.\label{eq:approx_rx_signal_X11_UL}
\end{equation}

\end{singlespace}

Note that unlike the discussion in subsection~\ref{sub:Ib}, it is
not accurate to further approximate $\hat{X}_{1}$ by a Gaussian RV,
because the randomness of the Gaussian distributed RV $S_{11}$ is
largely removed by the UL transmission power control mechanism, rendering
a less dominant role of the Gaussian distribution of $G_{1}$ compared
with the distribution of $H_{11}$. In other words, $\ensuremath{\sigma_{G_{1}}^{2}}$
is comparable with or even smaller than the variance of $H_{11}$,
making the approximation error large according to our results in~\cite{Our_DNA_work_TWC15}.
Therefore, we derive the approximate distribution of $X_{1}$ using
a different method, presented in Theorem~\ref{thm:Approx_CDF_X1}.
\begin{thm}
\label{thm:Approx_CDF_X1}The approximate CDF of $X_{1}$ is derived
as

\begin{singlespace}
\noindent
\begin{equation}
{F_{X_{1}}}\hspace{-0.1cm}\left(x\right)\hspace{-0.1cm}\approx\hspace{-0.1cm}{F_{\hat{X}_{1}}}\hspace{-0.1cm}\left(x\right)\hspace{-0.1cm}=\hspace{-0.1cm}\frac{1}{{\sqrt{\pi}}}\hspace{-0.1cm}\sum_{m=1}^{M_{0}}\hspace{-0.1cm}{w_{m}{F_{H_{11}}}\hspace{-0.1cm}\left(x\hspace{-0.1cm}-\hspace{-0.1cm}\left(\sqrt{2}\sigma_{G_{1}}a_{m}\hspace{-0.1cm}+\hspace{-0.1cm}\mu_{G_{1}}\right)\hspace{-0.1cm}\right)},\label{eq:thm_approx_CDF_X1}
\end{equation}

\end{singlespace}

\noindent where $M_{0}$ is the number of terms employed in the Gauss-Hermite
numerical integration~\cite{GH_num_integration}, and the weights
$\left\{ w_{m}\right\} $ and the abscissas $\left\{ a_{m}\right\} $
are tabulated in Table 25.10 of~\cite{GH_num_integration}. \end{thm}
\begin{IEEEproof}
Since $G_{1}$ is a Gaussian RV with the mean and the variance shown
in (\ref{eq:approx_signal_LxLN_mean_and_var}), the PDF of $G_{1}$
can be written as

\noindent
\begin{equation}
{f_{G_{1}}}\left(v\right)=\frac{1}{{\sqrt{2\pi\ensuremath{\sigma_{G_{1}}^{2}}}}}\exp\left\{ \hspace{-0.1cm}{-\frac{{\left({v-{\mu_{G_{1}}}}\right)^{2}}}{{2\ensuremath{\sigma_{G_{1}}^{2}}}}}\hspace{-0.1cm}\right\} .\label{eq:PDF_CDF_Gauss_G11}
\end{equation}

\begin{comment}
$\begin{cases}
\textrm{PDF of }G_{1}:\:\hspace{-0.1cm}\hspace{-0.1cm}{f_{G_{1}}}\left(v\right)=\frac{1}{{\sqrt{2\pi\ensuremath{\sigma_{G_{1}}^{2}}}}}\exp\left\{ \hspace{-0.1cm}{-\frac{{\left({v-{\mu_{G_{1}}}}\right)^{2}}}{{2\ensuremath{\sigma_{G_{1}}^{2}}}}}\hspace{-0.1cm}\right\}  & \hspace{-0.1cm}\hspace{-0.3cm}\\
\textrm{CDF of }G_{1}:\:\hspace{-0.1cm}\hspace{-0.1cm}{F_{G_{1}}}\left(v\right)=\Phi\left(\frac{v-{\mu_{G_{1}}}}{\ensuremath{\sigma_{G_{1}}}}\right) & \hspace{-0.1cm}\hspace{-0.3cm}
\end{cases}.$
\end{comment}

Besides, according to the definition of RVs in Section~\ref{sec:Network-Model},
we assume the CDF of $H_{11}$ to be ${F_{H_{11}}}\left(h\right)$.
Hence, the CDF of $X_{1}$ can be approximated by

\noindent ${F_{X_{1}}}\left(x\right)\approx{F_{\hat{X}_{1}}}\left(x\right)$

\begin{singlespace}
\noindent
\begin{eqnarray}
\hspace{-0.1cm}\hspace{-0.1cm}\hspace{-0.3cm} & = & \hspace{-0.3cm}\Pr\left[G_{1}+H_{11}\le x\right]\nonumber \\
\hspace{-0.1cm}\hspace{-0.1cm}\hspace{-0.3cm} & = & \hspace{-0.3cm}\Pr\left[H_{11}\le x-G_{1}\right]\nonumber \\
\hspace{-0.1cm}\hspace{-0.1cm}\hspace{-0.3cm} & = & \hspace{-0.3cm}\int\limits _{-\infty}^{+\infty}{{F_{H_{11}}}\left(x-v\right){f_{G_{1}}}\left(v\right)dv}\nonumber \\
\hspace{-0.1cm}\hspace{-0.1cm}\hspace{-0.3cm} & \overset{\left(a\right)}{=} & \hspace{-0.3cm}\int\limits _{-\infty}^{+\infty}{{F_{H_{11}}}\left(x-v\right)\frac{1}{{\sqrt{2\pi\ensuremath{\sigma_{G_{1}}^{2}}}}}\exp\left\{ {-\frac{{\left({v-{\mu_{G_{1}}}}\right)^{2}}}{{2\ensuremath{\sigma_{G_{1}}^{2}}}}}\right\} dv}\nonumber \\
\hspace{-0.1cm}\hspace{-0.1cm}\hspace{-0.3cm} & \overset{\left(b\right)}{=} & \hspace{-0.3cm}\frac{1}{{\sqrt{\pi}}}\int\limits _{-\infty}^{+\infty}{{F_{H_{11}}}\left(x-\left(\sqrt{2}\sigma_{G_{1}}y+\mu_{G_{1}}\right)\right)\exp\left(-y^{2}\right)dy}\nonumber \\
\hspace{-0.1cm}\hspace{-0.1cm}\hspace{-0.3cm} & \overset{\left(c\right)}{=} & \hspace{-0.3cm}\frac{1}{{\sqrt{\pi}}}\sum_{m=1}^{M_{0}}{w_{m}{F_{H_{11}}}\left(x-\left(\sqrt{2}\sigma_{G_{1}}a_{m}+\mu_{G_{1}}\right)\right)}+R_{M_{0}}\nonumber \\
\hspace{-0.1cm}\hspace{-0.1cm}\hspace{-0.3cm} & \overset{\left(d\right)}{\approx} & \hspace{-0.3cm}\frac{1}{{\sqrt{\pi}}}\sum_{m=1}^{M_{0}}{w_{m}{F_{H_{11}}}\left(x-\left(\sqrt{2}\sigma_{G_{1}}a_{m}+\mu_{G_{1}}\right)\right)},\hspace{-0.1cm}\hspace{-0.1cm}\hspace{-0.1cm}\hspace{-0.1cm}\hspace{-0.1cm}\hspace{-0.1cm}\hspace{-0.1cm}\hspace{-0.1cm}\label{eq:approx_CDF_X1}
\end{eqnarray}

\end{singlespace}

\noindent where the step (a) of (\ref{eq:approx_CDF_X1}) is obtained
from (\ref{eq:PDF_CDF_Gauss_G11}), and the step (b) of (\ref{eq:approx_CDF_X1})
is computed using the variable change $v=\sqrt{2}\sigma_{G_{1}}y+\mu_{G_{1}}$.
Moreover, the step (c) of (\ref{eq:approx_CDF_X1}) is derived using
the Gauss-Hermite numerical integration~\cite{GH_num_integration},
i.e., $\int\limits _{-\infty}^{+\infty}{f\left(y\right)\exp\left(-y^{2}\right)dy}=\sum_{m=1}^{M_{0}}{w_{m}f\left(a_{m}\right)}+R_{M_{0}}$,
where $M_{0}$ is the number of terms in the approximation, the weights
$\left\{ w_{m}\right\} $ and the abscissas $\left\{ a_{m}\right\} $
are tabulated in Table 25.10 of~\cite{GH_num_integration} and $R_{M_{0}}$
is a residual error in the order of $\ensuremath{\frac{{M_{0}!}}{{{2^{M_{0}}}\left({2M_{0}}\right)!}}}$~\cite{GH_num_integration},
which decays very fast as $M_{0}$ increases. %
\begin{comment}
\noindent The typical value of $M_{0}$ is 10 for practical use.
\end{comment}
Finally, the step (d) of (\ref{eq:approx_CDF_X1}) is obtained by
dropping $R_{M_{0}}$. Our proof is thus completed by comparing (\ref{eq:thm_approx_CDF_X1})
and (\ref{eq:approx_CDF_X1}).
\end{IEEEproof}
\begin{comment}
Placeholder
\end{comment}

In case of Rayleigh fading~\cite{Book_Proakis}, we propose Corollary~\ref{cor:approx_CDF_X1_Rayleigh}
to compute the approximate expression of ${F_{X_{1}}}\left(x\right)$.
\begin{cor}
\label{cor:approx_CDF_X1_Rayleigh}In case of Rayleigh fading, the
approximate CDF of $X_{1}$ can be computed by (\ref{eq:thm_approx_CDF_X1}),
where

\begin{singlespace}
\noindent
\begin{equation}
{F_{H_{11}}}\left(h\right)=1-\exp\left(-\exp\left(\frac{h}{\zeta}\right)\right),\label{eq:cor_CDF_H11_Rayleigh}
\end{equation}

\end{singlespace}

\noindent where $\zeta=\frac{10}{\ln10}$.\end{cor}
\begin{IEEEproof}
As discussed in Section~\ref{sec:Network-Model}, on condition of
Rayleigh fading, the channel gain $\left|{\bf {h}}_{11}\right|^{2}$
follows an exponential distribution with unitary mean~\cite{Book_Proakis}.
Then, our proof is completed by deriving (\ref{eq:cor_CDF_H11_Rayleigh})
based on the variable change $H_{11}=10\log_{10}\left|{\bf {h}}_{11}\right|^{2}$.
Details are omitted for brevity.
\end{IEEEproof}
\begin{comment}
Placeholder
\end{comment}

In case of Nakagami fading~\cite{Book_Proakis}, we propose Corollary~\ref{cor:approx_CDF_X1_Nakagami}
to compute the approximate expression of ${F_{X_{1}}}\left(x\right)$.
\begin{cor}
\label{cor:approx_CDF_X1_Nakagami}In case of Nakagami fading, the
approximate CDF of $X_{1}$ can be computed by (\ref{eq:thm_approx_CDF_X1}),
where

\begin{singlespace}
\noindent
\begin{equation}
{F_{H_{11}}}\left(h\right)=\frac{1}{\Gamma\left(k\right)}\gamma\left(k,\thinspace\frac{1}{\theta}\exp\left(\frac{h}{\zeta}\right)\right),\label{eq:cor_CDF_H11_Nakagami}
\end{equation}

\end{singlespace}

\noindent where $\Gamma\left(\cdot\right)$ and $\gamma\left(\cdot,\cdot\right)$
are respectively the gamma and the incomplete gamma functions~\cite{GH_num_integration},
$k$ and $\theta$ are respectively the shape and the scale parameters
of the Gamma distribution associated with the channel gain of Nakagami
fading~\cite{Book_Proakis}.\end{cor}
\begin{IEEEproof}
As discussed in Section~\ref{sec:Network-Model}, on condition of
Nakagami fading, the channel gain $\left|{\bf {h}}_{11}\right|^{2}$
follows a Gamma distribution with parameters $k$ and $\theta$~\cite{Book_Proakis}.
Then, our proof is completed by deriving (\ref{eq:cor_CDF_H11_Nakagami})
based on the variable change $H_{11}=10\log_{10}\left|{\bf {h}}_{11}\right|^{2}$
Details are omitted for brevity.
\end{IEEEproof}

\subsection{The SIR Analysis\label{sec:Analysis-of-SIR}}

\begin{comment}
Placeholder
\end{comment}

From (\ref{eq:rx_signal_X11_UL}), we can approximate the UL SIR in
dB by

\begin{singlespace}
\noindent
\begin{equation}
Z^{\textrm{dB}}\approx X_{1}-Q\overset{\triangle}{=}\hat{Z}^{\textrm{dB}}.\label{eq:approx_SIR_Z_dB}
\end{equation}

\end{singlespace}

\noindent We derive the approximate distribution of $Z^{\textrm{dB}}$
in Theorem~\ref{thm:approx_CDF_SIR_ZdB}.
\begin{thm}
\label{thm:approx_CDF_SIR_ZdB}The approximate CDF of $Z^{\textrm{dB}}$
is derived as

\noindent ${F_{Z^{\textrm{dB}}}}\left(z\right)\approx{F_{\hat{Z}^{\textrm{dB}}}}\left(z\right)$

\begin{singlespace}
\noindent
\begin{equation}
\hspace{-0.1cm}\hspace{-0.1cm}\hspace{-0.1cm}\hspace{-0.1cm}\hspace{-0.1cm}=\frac{\lambda}{{\sqrt{\pi}}}\hspace{-0.1cm}\sum_{m=1}^{M_{0}}\hspace{-0.1cm}{w_{m}\Phi^{\lambda-1}\hspace{-0.1cm}\left(\sqrt{2}a_{m}\right)\hspace{-0.1cm}{F_{X_{1}}}\hspace{-0.1cm}\left(z+\sqrt{2}\ensuremath{\sigma_{Q}}a_{m}+\mu_{Q}\right)},\hspace{-0.1cm}\hspace{-0.1cm}\hspace{-0.1cm}\label{eq:thm_approx_CDF_SIR_ZdB}
\end{equation}

\end{singlespace}

\noindent where $M_{0}$ is the number of terms employed in the Gauss-Hermite
numerical integration~\cite{GH_num_integration}, and the weights
$\left\{ w_{m}\right\} $ and the abscissas $\left\{ a_{m}\right\} $
are tabulated in Table 25.10 in~\cite{GH_num_integration}. \end{thm}
\begin{IEEEproof}
From (\ref{eq:approx_SIR_Z_dB}), the approximate CDF of $Z^{\textrm{dB}}$
can be derived as

\noindent ${F_{Z^{\textrm{dB}}}}\left(z\right)\approx{F_{\hat{Z}^{\textrm{dB}}}}\left(z\right)$

\begin{singlespace}
\noindent
\begin{eqnarray*}
\hspace{-0.3cm}\hspace{-0.1cm}\hspace{-0.1cm} & = & \hspace{-0.3cm}\Pr\left[X_{1}-Q\le z\right]\\
\hspace{-0.3cm}\hspace{-0.1cm}\hspace{-0.1cm} & = & \hspace{-0.3cm}\Pr\left[X_{1}\le z+Q\right]\\
\hspace{-0.3cm}\hspace{-0.1cm}\hspace{-0.1cm} & = & \hspace{-0.3cm}\hspace{-0.1cm}\hspace{-0.1cm}\int\limits _{-\infty}^{+\infty}\hspace{-0.1cm}\hspace{-0.1cm}{{F_{X_{1}}}\hspace{-0.1cm}\left(z+q\right){f_{Q}}\left(q\right)dq}\\
\hspace{-0.3cm}\hspace{-0.1cm}\hspace{-0.1cm} & \overset{\left(a\right)}{\approx} & \hspace{-0.3cm}\hspace{-0.1cm}\hspace{-0.1cm}\int\limits _{-\infty}^{+\infty}\hspace{-0.1cm}{{F_{\hat{X}_{1}}}\hspace{-0.1cm}\left(z+q\right)\lambda\Phi^{\lambda-1}\hspace{-0.1cm}\left(\hspace{-0.1cm}\frac{q\hspace{-0.1cm}-\hspace{-0.1cm}{\mu_{Q}}}{\sigma_{Q}}\hspace{-0.1cm}\right)\hspace{-0.1cm}\frac{1}{\sqrt{2\pi\ensuremath{\sigma_{Q}^{2}}}}\exp\left\{ {-\frac{{\left({q\hspace{-0.1cm}-\hspace{-0.1cm}{\mu_{Q}}}\right)^{2}}}{{2\ensuremath{\sigma_{Q}^{2}}}}}\hspace{-0.1cm}\right\} \hspace{-0.1cm}dq}\\
\hspace{-0.3cm}\hspace{-0.1cm}\hspace{-0.1cm} & \overset{\left(b\right)}{=} & \hspace{-0.3cm}\frac{1}{{\sqrt{\pi}}}\hspace{-0.1cm}\hspace{-0.1cm}\int\limits _{-\infty}^{+\infty}\hspace{-0.1cm}{{F_{\hat{X}_{1}}}\hspace{-0.1cm}\left(z+\sqrt{2}\ensuremath{\sigma_{Q}}y+\mu_{Q}\right)\hspace{-0.1cm}\lambda\Phi^{\lambda-1}\hspace{-0.1cm}\left(\sqrt{2}y\right)\exp\left(-y^{2}\right)\hspace{-0.1cm}dy}\\
\hspace{-0.3cm}\hspace{-0.1cm}\hspace{-0.1cm} & \overset{\left(c\right)}{=} & \hspace{-0.3cm}\frac{1}{{\sqrt{\pi}}}\hspace{-0.1cm}\sum_{m=1}^{M_{0}}\hspace{-0.1cm}{w_{m}{F_{\hat{X}_{1}}}\hspace{-0.1cm}\left(\hspace{-0.1cm}z+\sqrt{2}\ensuremath{\sigma_{Q}}a_{m}+\mu_{Q}\hspace{-0.1cm}\right)\hspace{-0.1cm}\lambda\Phi^{\lambda-1}\hspace{-0.1cm}\left(\sqrt{2}a_{m}\right)}\hspace{-0.1cm}+\hspace{-0.1cm}R_{M_{0}}
\end{eqnarray*}

\noindent
\begin{equation}
\hspace{-0.8cm}\hspace{-0.1cm}\overset{\left(d\right)}{\approx}\frac{\lambda}{{\sqrt{\pi}}}\hspace{-0.1cm}\sum_{m=1}^{M_{0}}\hspace{-0.1cm}{w_{m}\Phi^{\lambda-1}\hspace{-0.1cm}\left(\sqrt{2}a_{m}\right)\hspace{-0.1cm}{F_{\hat{X}_{1}}}\hspace{-0.1cm}\left(\hspace{-0.1cm}z+\sqrt{2}\ensuremath{\sigma_{Q}}a_{m}+\mu_{Q}\hspace{-0.1cm}\right)},\hspace{-0.1cm}\hspace{-0.1cm}\hspace{-0.1cm}\label{eq:approx_CDF_SIR_ZdB}
\end{equation}

\end{singlespace}

\noindent where the step (a) of (\ref{eq:approx_CDF_SIR_ZdB}) is
calculated using Theorem~\ref{thm:Approx_CDF_X1} and (\ref{eq:PDF_CDF_powerLN_Q}),
and the step (b) of (\ref{eq:approx_CDF_SIR_ZdB}) is computed using
the variable change $q=\sqrt{2}\sigma_{Q}y+\mu_{Q}$. Moreover, the
step (c) of (\ref{eq:approx_CDF_SIR_ZdB}) is derived using the Gauss-Hermite
numerical integration~\cite{GH_num_integration}. Finally, the step
(d) of (\ref{eq:approx_CDF_SIR_ZdB}) is obtained by dropping $R_{M_{0}}$.
Our proof is thus completed by comparing (\ref{eq:thm_approx_CDF_SIR_ZdB})
and (\ref{eq:approx_CDF_SIR_ZdB}).
\end{IEEEproof}
\begin{comment}
Placeholder
\end{comment}

In case of Rayleigh fading~\cite{Book_Proakis}, we propose Corollary~\ref{cor:approx_CDF_SIR_ZdB_Rayleigh}
to compute the approximate expression of ${F_{Z^{\textrm{dB}}}}\left(z\right)$.
\begin{cor}
\label{cor:approx_CDF_SIR_ZdB_Rayleigh}In case of Rayleigh fading,
the approximate CDF of $Z^{\textrm{dB}}$ can be computed by (\ref{eq:thm_approx_CDF_SIR_ZdB}),
where (\ref{eq:cor_CDF_H11_Rayleigh}) is plugged into (\ref{eq:thm_approx_CDF_X1})
to obtain ${F_{\hat{X}_{1}}}\left(x\right)$ in (\ref{eq:thm_approx_CDF_SIR_ZdB}). \end{cor}
\begin{IEEEproof}
The proof is completed by applying Corollary~\ref{cor:approx_CDF_X1_Rayleigh}
to Theorem~\ref{thm:Approx_CDF_X1} and Theorem~\ref{thm:approx_CDF_SIR_ZdB}.
Details are omitted for brevity.
\end{IEEEproof}
\begin{comment}
Placeholder
\end{comment}

In case of Nakagami fading~\cite{Book_Proakis}, we propose Corollary~\ref{cor:approx_CDF_SIR_ZdB_Nakagami}
to compute the approximate expression of ${F_{Z^{\textrm{dB}}}}\left(z\right)$.
\begin{cor}
\label{cor:approx_CDF_SIR_ZdB_Nakagami}In case of Nakagami fading,
the approximate CDF of $Z^{\textrm{dB}}$ can be computed by (\ref{eq:thm_approx_CDF_SIR_ZdB}),
where (\ref{eq:cor_CDF_H11_Nakagami}) is plugged into (\ref{eq:thm_approx_CDF_X1})
to obtain ${F_{\hat{X}_{1}}}\left(x\right)$ in (\ref{eq:thm_approx_CDF_SIR_ZdB}).\end{cor}
\begin{IEEEproof}
The proof is completed by applying Corollary~\ref{cor:approx_CDF_X1_Nakagami}
to Theorem~\ref{thm:Approx_CDF_X1} and Theorem~\ref{thm:approx_CDF_SIR_ZdB}.
Details are omitted for brevity.
\end{IEEEproof}

\section{Macroscopic Upgrade of the DNA-GA Analysis\label{sec:Macroscopic-Upgrade}}

\begin{comment}
Placeholder
\end{comment}

With Theorem~\ref{thm:approx_CDF_SIR_ZdB}, we have crafted a powerful
microscopic analysis tool based on the proposed DNA-GA analysis that
can deal with a wide range of network assumptions and system parameters.
In this section, we further investigate two approaches for upgrading
the DNA-GA analysis from a microscopic analysis tool to a macroscopic
one, putting DNA-GA in the same league as, e.g., stochastic geometry.

\subsection{The Semi-Analytical Approach\label{sub:The-Semi-Analytical-Approach}}

\begin{comment}
Placeholder
\end{comment}

The microscopic and the macroscopic analyses are closely related to
each other. The average performance of many microscopic analyses conducted
over a large number of random BS deployments converges to the performance
of the macroscopic analysis, given that the examined realizations
of the deterministic BS deployments follow the BS deployment assumption
used in the macroscopic analysis. Therefore, we can directly average
the performance results obtained by applying DNA-GA, i.e., Theorem~\ref{thm:approx_CDF_SIR_ZdB},
over a large number of random BS deployments to obtain the performance
results of the macroscopic analysis.

\subsection{The Analytical Approach\label{sub:The-Analytical-Approach}}

\begin{comment}
Placeholder
\end{comment}

Instead of conducting DNA-GA, i.e., Theorem~\ref{thm:approx_CDF_SIR_ZdB},
over many BS deployments and averaging all the results together to
obtain the results of the macroscopic analysis, we can construct an
idealistic BS deployment on a hexagonal lattice with \emph{the equivalent
BS density}, and perform a single DNA-GA analysis on such BS deployment
to extract an upper-bound of the SIR performance of the macroscopic
analysis. The hexagonal lattice leads to an upper-bound performance
because BSs are evenly distributed in the scenario and thus very strong
interference due to close proximity is avoided~{[}3,4{]}.

\section{Simulation and Discussion\label{sec:Simulaiton-and-Discussion}}

\begin{comment}
Placeholder
\end{comment}

In this section, we conduct simulations to validate the proposed DNA-GA
analysis, using both the semi-analytical and the analytical approaches.
For the semi-analytical approach, to obtain the results of the macroscopic
analysis, we average the results given by Theorem~\ref{thm:approx_CDF_SIR_ZdB}
over \emph{1000} random BS deployments. For each BS deployment, \emph{10,000}
random experiments are conducted to go through the randomness of UE
positions. And for each BS deployment and each UE placement, another\emph{
10,000} random experiments are conducted to go through the randomness
of shadow fading and multi-path fading. For the analytical approach,
only one BS deployment in a hexagonal lattice is examined. $M_{0}$
is set to 30 for the computation in the DNA-GA analysis to ensure
a good accuracy of the results~\cite{GH_num_integration}.

With regard to the scenario and parameters, 3rd Generation Partnership
Project (3GPP) recommendations have been considered~\cite{TR36.828}.
For the semi-analytical approach, 19 dummy macrocell sites are deployed
with a 0.5\ km inter-site distance to guide the small cell deployment.
Each macrocell site has the shape of a hexagon, and is equally divided
into 3 macrocells. Each macrocell contains 4 randomly deployed small
cells, resulting in $19\times3\times4=228$ small cells with a density
around 55.43\ $\textrm{cells/km}{}^{2}$. For the analytical approach,
the 228 small cells are located in a hexagonal lattice with the same
cell density. In both cases, each small cell has a coverage radius
of 0.04\ km, and the minimum inter-BS distance and the minimum BS-to-UE
distance are 0.04\ km, and 0.01\ km, respectively. Moreover, according
to~\cite{TR36.828}, $A=145.4$, $\alpha=3.75$, $P_{0}=-76$\ dBm,
$\eta=0.8$, and $\sigma_{S}=10$\ dB.

\begin{figure}[H]
\noindent \begin{centering}
\vspace{-0.1cm}
\includegraphics[width=6cm]{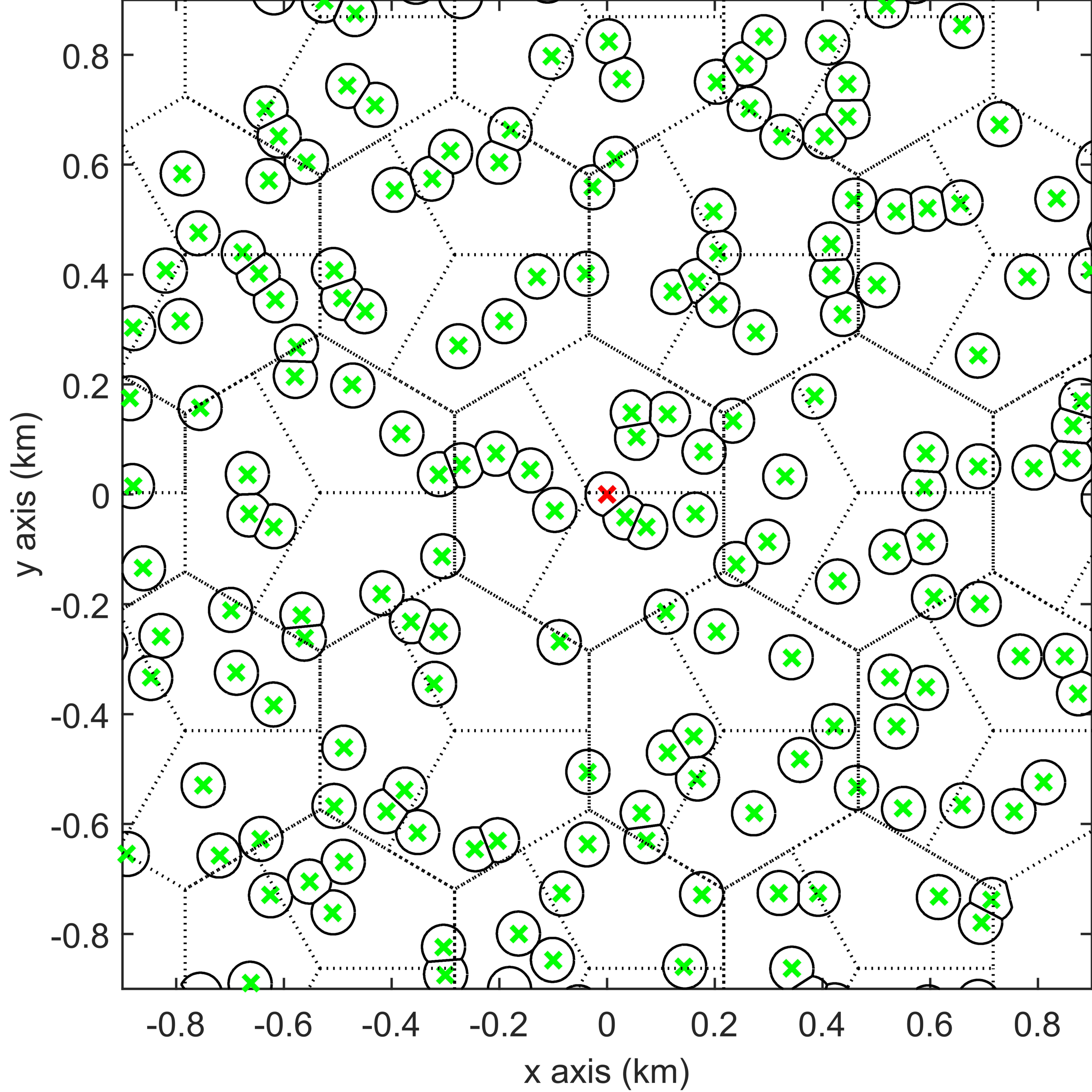}\renewcommand{\figurename}{Fig.}\protect\caption{\label{fig:3gpp_228cells_wMacro}Illustration of a 3GPP SCN deployment~\cite{TR36.828}.}

\par\end{centering}

\vspace{-0.3cm}
\end{figure}

Fig.~\ref{fig:3gpp_228cells_wMacro} illustrates an example of a
random BS deployment according to~\cite{TR36.828}, where small cell
BSs are represented by x-markers, while the coverage areas of dummy
macrocells and small cells are marked by dashed and solid lines, respectively.
UEs are randomly distributed in the mentioned small cell coverage
areas, and it is important to note that although some small cell coverage
areas are disk-shaped, the coverage areas of most small cells are
of irregular shape due to overlapping.

For brevity, in the following subsections, we omit the detailed investigation
on the interference analysis and the signal power analysis, and directly
present SIR results given by the DNA-GA analysis and the simulation.

\noindent \begin{center}
\vspace{-1cm}

\par\end{center}

\subsection{Validation of the DNA-GA Analysis}

\begin{comment}
Placeholder
\end{comment}

In this subsection, we validate the accuracy of the DNA-GA analysis
in terms of the SIR performance when assuming two cases for UE distribution
and multi-path fading, i.e.,
\begin{itemize}
\item Case 1: Uniform UE distribution + Rayleigh fading
\item Case 2: Non-uniform UE distribution + Nakagami fading
\end{itemize}
\begin{comment}
Placeholder
\end{comment}

For Case~1, we obtain the SIR results through the DNA-GA analysis
using Theorem~\ref{thm:approx_CDF_SIR_ZdB} and Corollary~\ref{cor:approx_CDF_SIR_ZdB_Rayleigh},
while for Case~2, we invoke Theorem~\ref{thm:approx_CDF_SIR_ZdB}
and Corollary~\ref{cor:approx_CDF_SIR_ZdB_Nakagami}.

When considering a non-uniform UE distribution, we assume that $f_{Z_{b}}\left(z\right)=\frac{W}{\rho},z\in R_{b}$,
where $\rho$ is the radial coordinate of $z$ in the polar coordinate
system, the origin of which is placed at the position of the BS of
$C_{b}$ and $W$ is a normalization constant to make $\int_{R_{b}}f_{Z_{b}}\left(z\right)dz=1$.
In the resulting non-uniform UE distribution, UEs are more likely
to locate in the close vicinity of the BS of $C_{b}$ than at the
cell-edge%
\footnote{Note that the considered $f_{Z_{b}}\left(z\right)$ is just an example
of the non-uniformly distributed UEs in $R_{b}$, which reflects a
reasonable network planning, where small cell BSs have been deployed
at the center of UE clusters. Other forms of $f_{Z_{b}}\left(z\right)$
can be considered in our DNA-GA analysis as well. %
}. When considering Nakagami fading, we assume that $k=10$ and $\theta=0.1$,
which corresponds to a multi-path fading with a strong line-of-sight
(LoS) component~\cite{Book_Proakis}.

\begin{figure}[H]
\noindent \begin{centering}
\vspace{-0.1cm}
\includegraphics[width=7.5cm]{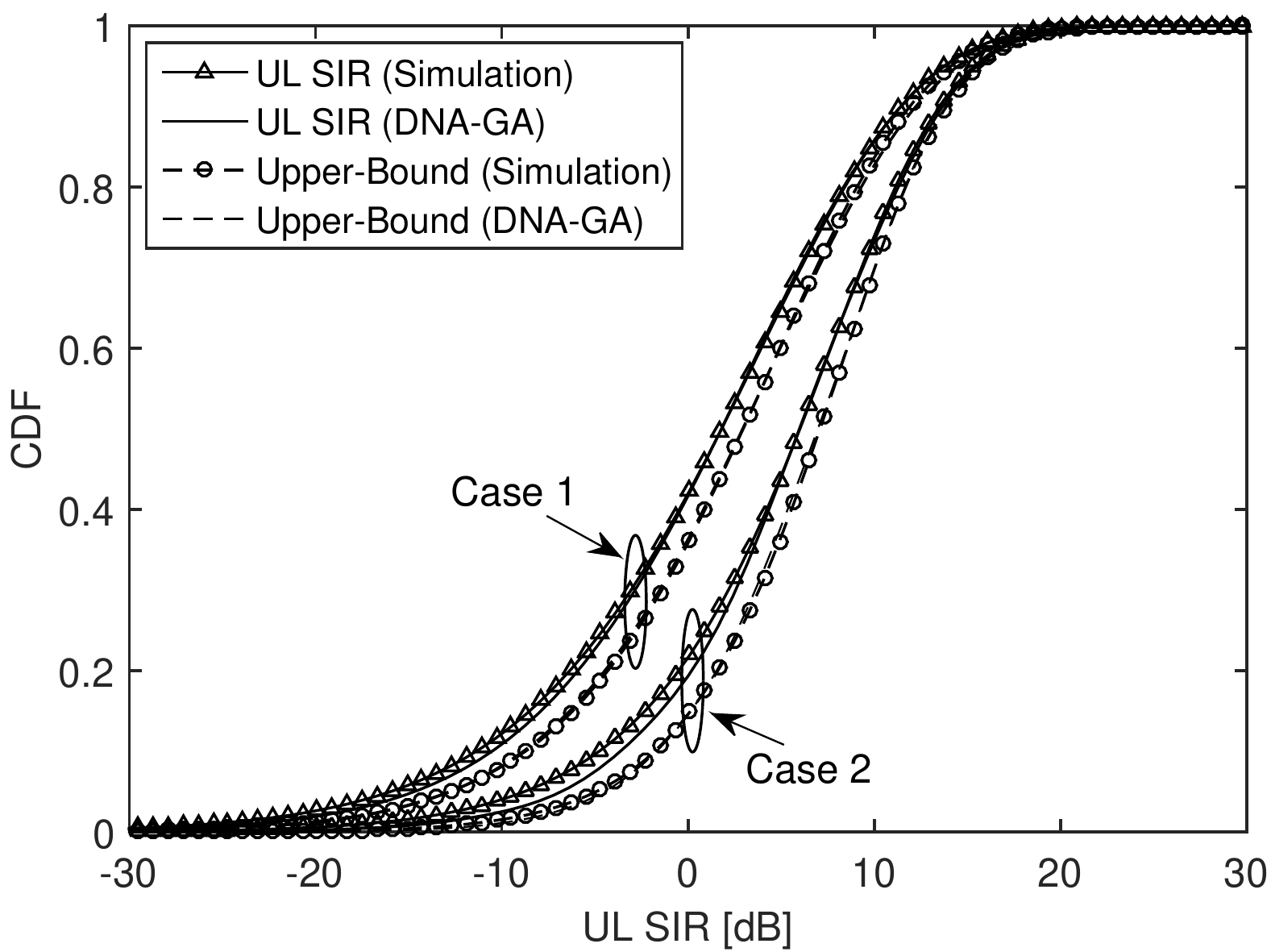}\renewcommand{\figurename}{Fig.}\protect\caption{\label{fig:3gpp_228cells_SIR_dB_DNA-GA}UL SIR in dB (DNA-GA vs. Simulation).}

\par\end{centering}

\vspace{-0.3cm}
\end{figure}

For both cases, the UL SIR performance is evaluated using the simulation
and the semi-analytical approach discussed in Subsection~\ref{sub:The-Semi-Analytical-Approach}.
Moreover, the upper bound of the UL SIR is also investigated using
the simulation and the analytical approach discussed in Subsection~\ref{sub:The-Analytical-Approach}
based on a BS deployment in a hexagonal lattice. The results are shown
in Fig.~\ref{fig:3gpp_228cells_SIR_dB_DNA-GA}.

\noindent \begin{center}
\vspace{-1cm}

\par\end{center}

\begin{comment}
The numerical results of $\lambda$, $\mu_{Q}$ and $\sigma_{Q}^{2}$
are provided
\end{comment}

As can be seen from Fig.~\ref{fig:3gpp_228cells_SIR_dB_DNA-GA},
the SIR results of the proposed DNA-GA analysis match those of the
simulation very well, particularly in the head portion. In the semi-analytical
approach, the maximum deviation between the CDFs obtained by the DNA-GA
analysis and the simulation for both investigated cases are around
$1.0\hspace{-0.1cm}\sim\hspace{-0.1cm}1.7$\,percentile. In the analytical
approach, the fitness becomes even better, i.e., the maximum deviation
for both cases is within $0.6$\,percentile.

More importantly, for both cases, the upper-bound SIR performance
given by the analytical approach is shown to be within $2.0\hspace{-0.1cm}\sim\hspace{-0.1cm}2.5$\,\,dB
from the exact performance, indicating its usefulness in characterizing
the network performance with low-complexity computation. To take Case~1
of DNA-GA as an example, the numerical results to be plugged into
Theorem~\ref{thm:approx_CDF_SIR_ZdB} for the hexagonal BS deployment
are $\mu_{G_{1}}=-93.07$, $\sigma_{G_{1}}^{2}=5.97$, $\lambda=202.66$,
$\mu_{Q}=-137.71$ and $\sigma_{Q}^{2}=212.04$.

Finally, note that the SIR of Case~2 outperforms that of Case~1,
mainly because UEs tend to stay closer to their serving BSs in Case~2
as discussed above, leading to a larger signal power and a lower interference
power.

\subsection{Comparison Between DNA-GA and Stochastic Geometry}

\begin{comment}
Placeholder
\end{comment}

\begin{figure}[H]
\noindent \begin{centering}
\vspace{-0.1cm}
\includegraphics[width=7.5cm]{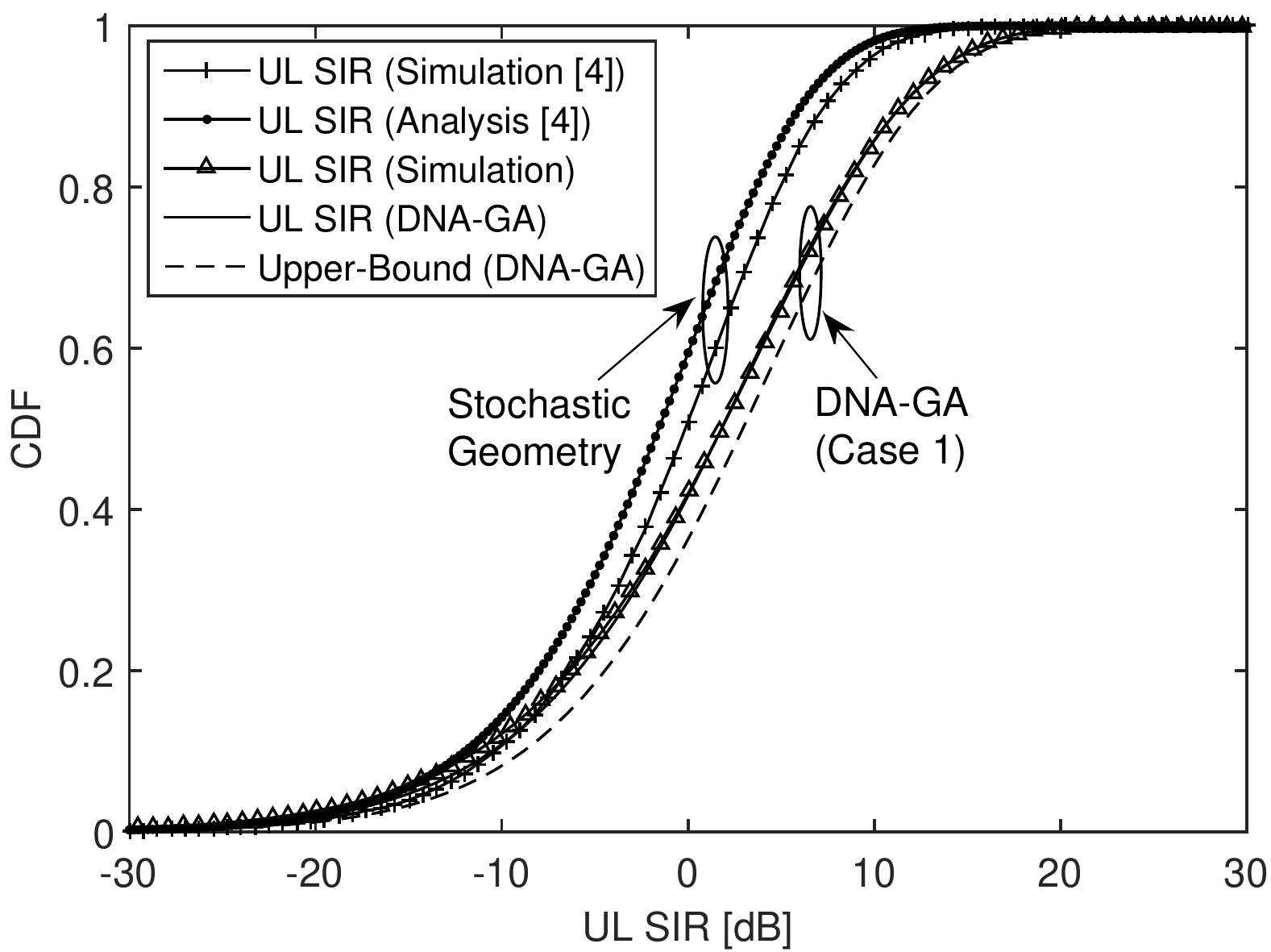}\renewcommand{\figurename}{Fig.}\protect\caption{\label{fig:DNA-GA_vs_HPPP}UL SIR in dB (DNA-GA vs. Stochastic Geometry~\cite{Jeff_UL}).}

\par\end{centering}

\vspace{-0.3cm}
\end{figure}

In this section, we compare the UL SIR results of the DNA-GA analysis
(Case~1) and those of the stochastic geometry analysis in Fig.~\ref{fig:DNA-GA_vs_HPPP},
with the same average cell density of 55.43\ $\textrm{cells/km}{}^{2}$
and the same assumption of Rayleigh fading%
\footnote{Note that the stochastic geometry analysis in~\cite{Jeff_UL} poses
some assumptions on the model for the sake of tractability, e.g.,
no shadow fading, Rayleigh fading only, homogeneous Poisson distribution
of both UEs and BSs in the entire scenario. In contrast, the DNA-GA
analysis does not need such assumptions and works with a more realistic
model, considering the hotspot SCN scenario~\cite{TR36.828} shown
in Fig.~\ref{fig:3gpp_228cells_wMacro} and discussed in Remark~2
of Section~\ref{sec:Network-Model}.%
}.

\noindent \begin{center}
\vspace{-1cm}

\par\end{center}

In Fig.~\ref{fig:DNA-GA_vs_HPPP}, a few interesting aspects are
noteworthy. First, both our analysis and the analysis in~\cite{Jeff_UL}
are only able to give approximate results. However, the approximation
error of our DNA-GA analysis is shown to be smaller than that of~\cite{Jeff_UL}%
\begin{comment}
 which indicates better accuracy of DNA-GA
\end{comment}
.

Second, there is a significant performance gap between our DNA-GA
analysis and the stochastic geometry analysis in~\cite{Jeff_UL}.
This is because (i) Our DNA-GA analysis considers the shadow fading
on top of the multi-path fading, which leads to a large variance of
SIR, while the analysis in~\cite{Jeff_UL} does not, which gives
a small SIR variance%
\begin{comment}
(the impossibility of considering shadowing within stochastic geometry
is one of its main drawbacks, which we overcome with DNA-GA) a larger
variance of SIR distribution
\end{comment}
; and (ii) The DNA-GA analysis studies the hotspot SCN scenario recommended
by the 3GPP~\cite{TR36.828}, where UEs are deployed closer to the
serving BSs than those in Voronoi cells considered in~\cite{Jeff_UL}.

Third, the purpose of Fig.~\ref{fig:DNA-GA_vs_HPPP} is not to reproduce
the results in~\cite{Jeff_UL} based on Voronoi cells, but to analytically
investigate a more practical 3GPP network scenario. If the shadow
fading is required to be ignored, albeit impractical, the Gamma approximation
of the aggregate interference~\cite{UL_interf_sim3} could be invoked
to make our approach of analysis still valid. Besides, the DNA-GA
analysis can also handle the case where the cell coverage areas are
constructed as Voronoi cells~\cite{Jeff_UL}. However, to do so,
it would be more practical to consider an alternative UE association
strategy (UAS), where each UE is connected to the BS with the smallest
path loss plus shadow fading. Note that such UAS will blur the boundaries
of Voronoi cells, because each UE is no longer always connected to
its closest BS, making the analysis more intricate and realistic.

Finally, note that a three-fold integral computation is needed in~\cite{Jeff_UL}
to compute the results, while no integration is required in Theorem~\ref{thm:approx_CDF_SIR_ZdB}
of the DNA-GA analysis%
\begin{comment}
, resulting in a lower complexity
\end{comment}
. However, many BS deployments are needed in the semi-analytical approach
of the DNA-GA analysis, while only one in the analytical approach.

\begin{comment}
Placeholder
\end{comment}
\begin{comment}
An important note on the proposed network performance analysis is
that our analytical approach is able to obtain the results in an efficient
manner.
\end{comment}

\begin{comment}
The computational complexity of the proposed approach is mainly attributable
to the numerical integration required to obtain the values of $\varepsilon$
for each small cell. In contrast, the simulation approach, e.g.,~\cite{UL_interf_sim1},~\cite{UL_interf_sim2},
as well as in this paper, involves a tremendously high complexity.
Specifically, in our simulations, in order to go through the randomness
of all the RVs discussed in Section~\ref{sec:Network-Model}, more
than \emph{one billion} of realizations of $I_{b}$ have been conducted
for the 83 interfering cells depicted in Fig.~\ref{fig:3gpp_84cells}.
This shows that the proposed microscopic analysis of network performance
is computationally efficient, which makes it a convenient tool to
study future 5G systems with general and dense small cell deployments.
Furthermore, our analytical studies yield better insight into the
performance of the system compared with simulations.
\end{comment}

\begin{comment}
\balance
\end{comment}

\section{Conclusion\label{sec:Conclusion}}

We proposed a new approach of network performance analysis, which
unifies the microscopic and the macroscopic analyses within a single
framework. Compared with stochastic geometry, our DNA-GA analysis
considers shadow fading, general UE distribution and any type of multi-path
fading, as well as any shape and/or size of cell coverage areas. Thus,
DNA-GA can analyze more realistic networks and is very useful for
the network performance analysis of the 5G systems with general cell
deployment and UE distribution.

\end{document}